\documentclass{article}
\usepackage{amssymb}
\usepackage{amsmath}
\usepackage{dcolumn}
\usepackage{bm}
\usepackage{caption}
\usepackage{graphicx}
\graphicspath{/home/kanchan/carmi}
\usepackage[utf8]{inputenc}
\usepackage[a4paper,total={6in,8in}]{geometry}
\usepackage[a4paper]{geometry}
\usepackage[arrowdel]{physics}
\begin{document}

\centerline{\LARGE {\bf Time reversed states in barrier tuneling}}
\centerline{ Kanchan Meena, P. Singha Deo}
\centerline{\bf S. N. Bose National Center for Basic Sciences, Salt Lake, Kolkata, India 700106.}

\begin{abstract}
Tunneling, though a physical reality, is shrouded in mystery.
Wave packets cannot be constructed under the barrier and group velocity cannot be defined. 
The tunneling particle can be observed on either sides of the barrier but its properties under the barrier has never been
probed due to several problems related to quantum measurement. We show that there are ways to bypass these problems in mesoscopic systems and one can even derive 
an expression for the quantum mechanical current under the barrier. 
A general scheme is developed to derive this expression for any arbitrary system.
One can use mesoscopic phenomenon to 
subject the expression to several theoretical and
experimental cross checks.
For demonstration we consider an ideal 1D quantum ring with Aharonov-Bohm flux $\Phi$, connected to a reservoir. 
It gives clear evidence that propagation occur under the barrier resulting in a current that can be measured non-invasively
and theoretically cross checked.
Time reversed states play a role but there is no evidence of violation of causality.
The evanescent states are known to be largely stable and robust against phase fluctuations making them a possible candidate for device applications
and so formalizing current under barrier is important.

\end{abstract}

\section{Introduction}

Evanescent modes are one of the puzzling features of quantum mechanics wherein a particle with energy less than the barrier height can tunnel through the barrier and
this has no classical counterpart. Although mathematical analogies can be found with electromagnetic wave-packets, physical interpretations pose a serious challenge.
It is not possible to draw any classical correspondence because one cannot construct a wave-packet with such evanescent states and
so one cannot define propagation of the particle under the barrier. This has lead to study several aspects of states under the barrier and one such aspect is tunneling
time. Since without a wave-packet one cannot define a group velocity and therefore addressing the problem lead to several puzzles
\cite{lan1}. In a situation where a
proper theory or formalism is missing making sense of experiments is difficult \cite{exp1}.
One can address the problem with the help of physical clocks like Larmor clock and Wigner delay time but that invariably
appeals to semi-classical concepts like spin precision and stationary phase approximations \cite{scl}.
Feynman path approach has also been applied \cite{feyn} and different approaches
give different results.
Hartman effect like phenomenon \cite{hart}
show that a tunneling particle can come out of the barrier before entering it which raises questions like if at all
there is propagation under the barrier.
We describe here a situation wherein one can make a theoretical experiment to confirm current due to evanescent modes and hence conclude propagation
under the barrier. That will also be a testing ground for the validity of physical clock like the Larmor clock. Of course one way to validate Larmor clock
is to appeal to Berger's circuit and analyticity \cite{deo}. There is no harm in making an alternate validation by showing that current
under the barrier as a measurable quantity can be established from analyticity of scattering matrix elements
even if one can not prove its existence from quantum mechanics alone. For this we will use the set up described in the next section.
For the complexity of physical interpretations we limit ourselves to the simplest system of
1D rings. 
The sample in the form of a
1D quantum ring is not a loss of generality.
Such 1D rings can be achieved experimentally using lateral
confinement and materials for which effective mass approximation works \cite{dat}.
Besides, as shown in the book by S. Datta \cite{mail}, one can make finite thickness rings that consist of many independent 1D channels.
A finite thickness lead and
ring can have multiple channels but DOS per channel is just the
1D DOS that reflects in conductance measurements \cite{dat}.

\section{The set up}

We will first provide a simple description of the mesoscopic
set up and then discuss how this set up helps us
address the problems discussed in the previous section.
The mesoscopic sample is taken to be a ring pierced by a magnetic flux $\Phi$ through the center of the ring such that there is no magnetic field on the
electrons confined to the ring.
The ring is usually made up of gold or copper or semiconductors that can accommodate an electron gas.
We are interested in the equilibrium
response of this sample which in this case is response to a magnetic field.
\begin{figure}[h!]
\centering
\includegraphics[scale=0.15]{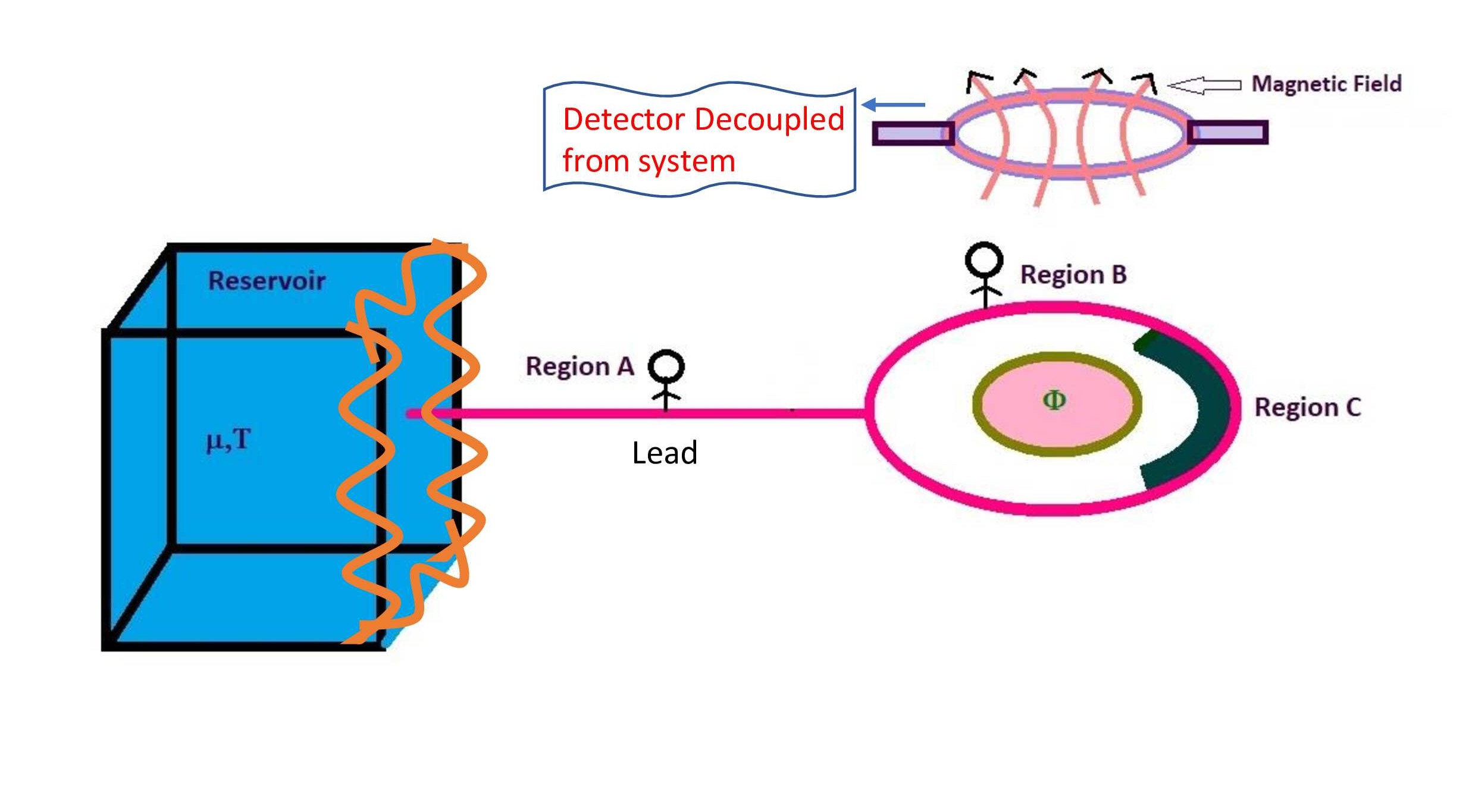}
\captionsetup{labelformat=empty}
\caption{Fig. 1 The set up we want to consider in this work is shown which is a typical mesoscopic grand canonical
system in equilibrium and described in detail in section 2.}
\end{figure}
There is a reservoir shown as a 3D block which is a source of electrons at chemical potential $\mu$ and temperature $T$,
making it a grand canonical system.
The reservoir injects electrons to the sample or to the ring through a lead shown as region A.
The ring can thus exchange electrons with the reservoir through the lead.
This exchange does not result in a net current
in the lead because electrons that go into the ring also come out of it and escape to the classical reservoir which is very typical of a voltage probe.
The system therefore constitutes an equilibrium system where the reservoir also acts as a source of
decoherence according to the Landauer-Buttiker formalism.

The electrons in the lead and
inside the ring are purely described by quantum mechanics. 
If the ring is isolated then it will have some eigen functions and energy eigen states that
can be obtained from Schrodinger equation. But once connected to the reservoir, the states in the ring will be affected. They will no longer be eigen-states of the
Hamiltonian but the eigen-states will acquire some broadening due to life time related effects as the state can now leak into the reservoir. The states
in the lead is unaffected by the ring because it is an ideal 1D system with a typical DOS given by $\frac{2 \pi m}{h^2k}$.
A part of the ring (region C in Fig. 1) has a potential $U$
that is shown as a thickened line. $U$ can be so adjusted that an electron can tunnel through this region C.
This region C can be extended to the entire ring wherein the entire
ring can be made into a tunneling region. We know the magnetic flux can drive an equilibrium persistent
current \cite{deo} in the ring and this current can thus be a tunneling current in
part of the ring or in the entire ring. This allows us to extend the theoretical
consistency of quantum currents to the tunneling regime.
We have shown in Eq. 13 that one has to follow a different approach under the barrier using analytic
continuity that does not lead to a straightforward
probabilistic interpretation of quantum mechanical currents.
When the barrier is located only in a part of the ring then the persistent current in the
ring can flow in a quantum state that is partly in the propagating regime (for example in the region B in Fig. 1)
and partly in the evanescent regime (for example in the region C in Fig. 1). Consistency between them has to conserve the current at
the point of propagating to evanescent crossover at the junction between regions B and C, there being no puzzle about the current in region B.

The primary advantage of theoretically studying such mesoscopic tunneling currents is that there are
alternate theoretical formalisms in mesoscopic regime
to verify the existence of currents inside the system under the barrier from propagating asymptotic states far from the
barrier. That means there are theoretical cross checks for propagation under the barrier without studying the states under the barrier.
Thus observations and calculation on current under the barrier can help us support or eliminate ideas.
In short there are asymptotic states for the current carrying states in the ring and one can determine the current in the ring from these asymptotic
states \cite{melakk}, that is again similar to the theory of the Larmor clock \cite{deo}.
Note that the derivation by Larmor clock \cite{deo} giving the LHS of Eq. 18 does not need Schrodinger equation in particular.
Now to calculate a scattering matrix element we may need an equation of motion but the derivation of LHS of Eq. 18 is insensitive to details like
whether one is allowed to analytically continue the wave-function above the barrier to below the barrier or one should have a separate
equation of motion for evanescent states. Rather one may interpret that since all that Eq. 16 and LHS of Eq. 18
cares is analyticity of scattering matrix elements,
it should be legitimate to analytically continue wave-functions inside the system as long as it does not destroy the analyticity of
scattering matrix elements. It has been claimed \cite{deo} that the Larmor clock is more fundamental than Schrodinger equation as it gives a lot of
new measurable quantities like the hierarchy of DOS that Schrodinger equation does not. So if there are doubts about what one gets from
Schrodinger equation, we may expect to get confirmation from the theory of Larmor clocks.
In the limit of a closed system it is also consistent with what one gets
from Schrodinger equation.
One can make the entire ring to be in the evanescent regime and the
currents due to evanescent states has to be consistent with the asymptotic theory.
A cartoon observer is shown in region A or lead of Fig. 1 and another cartoon observer is shown inside the ring.
Observer A need not have any idea about the potential inside the ring. This observer does not know if the tunneling region extend over the entire
ring or only a part of the ring. This observer only needs to know the infinitesimal change of $dU=\epsilon$ without any knowledge of $U$.
He can measure the scattering phase shift (a theoretical measurement using analysis)
in the wave-function in the lead and from there infer the current inside the ring from the analyticity of the scattering matrix elements.
Observer A can vary incident energy and check analiticity of scattering matrix elements from Cauchy-Riemann conditions or similar criterion.
The observer B in the ring can extend or shrink the tunneling region and also
determine the tunneling current from the analytic continuation of the internal wave-function which is an unsettled issue in quantum mechanics.
If the two measurements agree then the measurement by observer in the ring has to
be correct. 

In terms of practical measurements,
there are problems associated with measuring
tunneling currents in quantum mechanics and whatever we know about quantum measurements through an entanglement of sample states
with the states of the detector do not apply for evanescent modes. One can also not measure tunneling currents classically as that will require the detector
to be placed under the barrier and classical detectors can not work under the barrier. These problems can be avoided in the
above described mesoscopic set up.
Because such currents can also cause magnetization that can be observed in a practical experiment for further validation.
This magnetization can be measured without disturbing the state in the system and thus not invoking the unresolved issues of quantum measurement.
In mesoscopic systems electrons are quantum mechanical but fields are classical. We can measure the
magnetization due to the currents classically using hall magnetometers or squids remotely.
This is why Fig. 1 shows a detector to measure magnetization, placed above the ring.
Thus currents due to evanescent modes inside a ring can be measured
non-invasively without encountering the above mentioned problems of quantum measurement.
So it makes a lot of sense to study persistent currents due to evanescent modes inside the ring in the set up of
Fig. 1.

\section{Theoretical treatment}

\begin{figure}[h!]
\centering
\includegraphics[scale=0.55]{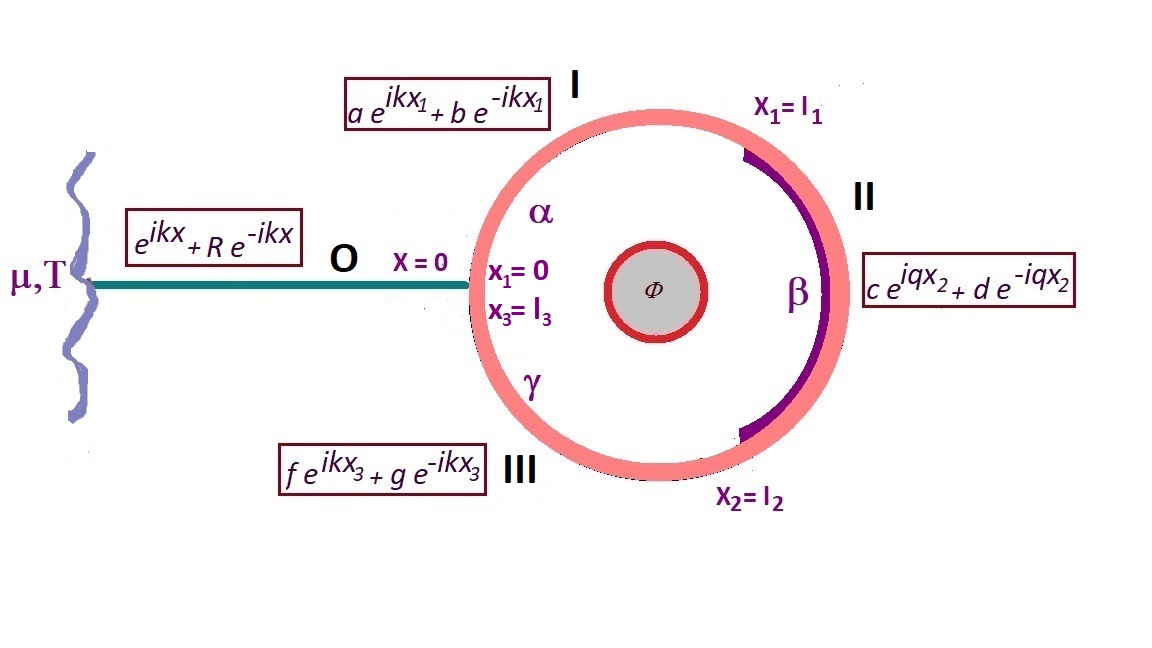}
\captionsetup{labelformat=empty}
\caption{Fig. 2 The same ring as in Fig. 1 is depicted here to show the different parameters used in the theory.
The ring region I is of length $l_1$, region II is of length $l_2$, and region III is of length $l_3$. The total length of the ring is
$L=l_1+l_2+l_3$. The choice of coordinates with their
origin is depicted in the figure. $\alpha$ is the Aharonov-Bohm (AB) phase for traversing region I, $\beta$ is that for region II and $\gamma$ is that for region III,
that appear in boundary conditions although the observable quantities like current can only depend on $\alpha+\beta+\gamma=\frac{2 \pi \Phi}{\Phi_0}= \Phi'$ (say).}
\end{figure}

In Fig. 2 we redraw the system in Fig. 1 in order to depict the parameters used and the wave-functions in the different regions for which the details
can be seen in the figure caption. Here the reservoir is shown as an irregular thick line on the left. Region A of Fig. 1 is called region O and the
three regions in the ring are labeled I, II and III.
The Schrodinger equation in 1D is
$$\frac{-\hbar^2}{2m} \frac{d^2\psi(x)}{dx^2} + U(x) \psi(x) = E \psi(x) $$
The solutions of Schrodinger equation in regions O (or lead), I, II and III of Fig. 2 give the wave functions in absence of
magnetic field as
\begin{equation}
\psi_0=\frac{1}{\sqrt k}(e^{ikx} + Re^{-ikx})
\end{equation}
\begin{equation}
\psi_{I}=\frac{1}{\sqrt k}(ae^{ikx_1} + be^{-ikx_1})
\end{equation}
\begin{equation}
\psi_{II}=\frac{1}{\sqrt q}(ce^{iqx_2} + de^{-iqx_2})
\end{equation}
\begin{equation}
\psi_{III}=\frac{1}{\sqrt k}(fe^{ikx_3} + ge^{-ikx_3})
\end{equation}
Here
\begin{equation}
q=(\frac{2m}{\hbar^2}(E-U))^{1/2} \;\; and \;\; k=(\frac{2m}{\hbar^2}E)^{1/2}
\end{equation}
It can be shown \cite{deo} that the magnetic field can be included through the boundary conditions and the magnetic field results in a current
in the ring.

The objective of this study is to consider the situation when $E<U$ in Eq. 5 and 3. Then the persistent current in region II is carried by evanescent modes.
In this situation $q \rightarrow is$ which is known as analytical continuation
and the expression for the current can be derived using the following scheme.
Although, such an analytic continuation completely destroyes the idea of a propagating wavepacket as a Hilbert space element, an analytic continuation
will always satisfy the Schrodinger equation.
First of all we substitute $ q \longrightarrow i s $ where $s=(\frac{2m}{\hbar^2}(U-E))^{1/2}$ in Eq. 3 and 5.
$ \psi_0 $, $ \psi_{I} $ , $\psi_{III}$ remain unchanged while $ \psi_{II} $ becomes
\begin{equation}
\psi_{II}=\frac{1}{\sqrt{s}} (c e^{i (is) x_2} + d e^{-i (is) x_2})
\end{equation}

Single valuedness of wavefunction implies
$$ \psi_0|_{x=0^-} = \psi_I|_{x_1 = 0^+} = \psi_{III}|_{x_3=l_{3}^-} $$
$$ \psi_I|_{x_1=l_{1}^-} = \psi_{II}|_{x_2=0^+} $$
\begin{equation}
\psi_{II}|_{x_2=l_{2}^-} = \psi_{III}|_{x_3=0^+}
\end{equation}
Also from current conservation or Kirchoff's law in one dimension one can write
at the junctions $x=0$, $x_1=l_1$ and $x_2=l_2$ (see Fig. 2)
\begin{equation}
\sum_i \frac{d\psi_i}{dx_i} = 0
\end{equation}
Quantum mechanical current is given by the expression
\begin{equation}
J(k,\Phi)=\frac{e\hbar}{2mi} \bigg[\psi^* \nabla \psi- HC \bigg]
\end{equation}

At the junction of regions I and II,
current conservation implies that the sum of currents arriving and leaving the points given by $x_1=l_1^-$ and $x_2=0^+$ is zero.
Thus Eq. 9 implies (ignoring the $\frac{e \hbar}{2mi}$ factor that can be added later)
\begin{equation}
\bigg[\psi_I^* \nabla \psi_I- HC \bigg]_{x_1=l_{1}^-}
- \bigg[\psi_{II}^* \nabla \psi_{II}- HC \bigg]_{x_2=0^{+}}=0
\end{equation}
From Eqs. 2 and 6
$$\frac{1}{k} \Bigg[(ae^{ikx_1 + i\Phi'} + b e^{-ikx_1})^* \frac{d}{dx_1}(ae^{ikx_1 + i\Phi'} + b e^{-ikx_1}) -HC\Bigg]_{x_1=l_{1}^-}$$
\begin{equation}
- \frac{1}{s} \Bigg[(c e^{i(is)x_2} + d e^{-i(is)x_2 - i \Phi'})^* \frac{d}{dx_2} (ce^{i(is)x_2} + d e^{-i(is)x_2 - i \Phi'}) -HC \Bigg]_{x_2=0^+}= 0
\end{equation}
Here $\Phi'=\frac{2 \pi \Phi}{\Phi_0}$ and
the fact that only some terms aquire this phase factor to account for the magnetic field
through the boundary conditions has been explained in detail using Feynman path approach in Ref. \cite{deo}.
From Eq. 11 it is clear that the first bracketed term give the current in region I and the second bracketed term give the current in region II.
Simplifying the second bracket along with its Hermitian conjugate and reintroducing the $\frac{e \hbar}{2mi}$ factor we get
\begin{equation}
- \frac{e \hbar}{2mi} \bigg[\mid{c}\mid^2 - c^* d e^{-i\Phi'} + d^* c e^{i \Phi'} - \mid{d}\mid^2 -HC \bigg]
\end{equation}
Subtracting the Hermitian conjugate we get current density expression in the region II as
\begin{equation}
J(k,\Phi)=\frac{e\hbar}{m i}\bigg[d c^* e^{-i \Phi'} - c d^* e^{i \Phi'} \bigg] 
\end{equation}
Differential current in an energy interval $dE$ will be
\begin{eqnarray}
dJ(k,\Phi)=\frac{e\hbar}{m i}\bigg[d c^* e^{-i \Phi'} - c d^* e^{i \Phi'} \bigg] \frac{dn}{dE} dE \nonumber \\
=\frac{e}{hi}\bigg[d c^* e^{-i \Phi'} - c d^* e^{i \Phi'} \bigg] dE
\end{eqnarray}
Similarly, current in region I where there is no potential, will be
\begin{equation}
dJ(k,\Phi)= \frac{e}{h} \bigg[|a|^2 - |b|^2 \bigg] dE
\end{equation}
Thus if $E>U$ then the current in the regions I, II and III will be given by
$\frac{e}{h} (|a|^2 -|b|^2) $, $\frac{e}{h} (|c|^2 - |d|^2) $ and
$\frac{e}{h}(|f|^2-|g|^2)$, respectively, and this follows from the standard definition of quantum mechanical current even in the presence of flux $\Phi$.
Continuity of current implies that they should be all equal and that obviously comes out in our calculations.
In absence of flux a, c and f are the amplitude of clockwise moving electrons while b, d and g are the amplitudes of anticlockwise electrons and they can only
differ by a relative phase because none of them is preferred over the other. Flux breaks this symmetry so that
$(|a|^2 -|b|^2)$ is the difference between the number of electrons moving clockwise
and those moving anticlockwise in region I and that constitutes a current.
This means that when $\Phi$ is zero then there is no current in the system as is expected for an equilibrium
situation. However for $\Phi \ne 0$ the current is finite. It is an equilibrium current called persistent
current, purely quantum mechanical in origin \cite{deo}.
Eq. 14 is far more complicated to be interpreted physically. Simple decoupling in terms of clockwise moving electron probability
and anticlockwise moving electron probability is not possible. Time reversed amplitudes $c^*$ and $d^*$
appear in the expression. Tunneling under the barrier has long been
studied as an avenue for superluminality but conclusive results do not exist as discussed in the introduction.
Current expression with current conservation conclusively show that time reversed states have a role
to play in the propagation.

This current is therefore
flowing in the ring and magnetizing the ring. This magnetization can be measured by the detector on top and hence can be also experimentally verified.
Without resorting to a practical experiment, this current expression for evanescent modes can be subjected to several tests. First is that it satisfies
current continuity at the junction of regions I and II (see Fig. 2).
Secondly an experimentalist can non-invasively measure and verify the numerical values of the current.
As a third verification consider theoretically reducing the regions I and III to zero lengths, in which case the current and magnetization
will be entirely due to evanescent modes.
Regions I and III being reduced to zero any magnetization observed has to be purely due to current carried by evanescent modes.
Such a magnetization of an entire ring carrying current due to tunneling will prove propagation under the barrier.
Without waiting for an experimentalist to measure and verify this there can be an independent theoretical verification which works only in the mesoscopic
regime as has been elaborated in section 2.
An observer in the lead can remotely calculate this current without even knowing if the regions I and III are reduced to zero length or they have a finite length.
For that we have a novel expression for current derived in [P. Mello, 1993] and [E. Akkermans et al, 1991] and stated below.
\begin{equation}
dJ(E,\Phi) = \frac{c}{\pi} \frac{\partial \theta_R}{\partial \Phi} dE = \frac{e}{h} \frac{\partial \theta_R}{\partial \Phi'} dE
\end{equation}
Here $dJ(E, \Phi)$ is the differential current in an energy interval $dE$ at a fixed flux $\Phi$ and $\Phi'=\frac{2\pi\Phi}{\Phi_0}$. The reflection phase shift
is defined as $\theta_R=Arctan[\frac{Im(R)}{Real(R)}]$.
The scattering phase shift
$\theta_R(E, \Phi)$ is therefore from the asymptotic wave-function in the lead far away from the ring.
Observable current will be
\begin{equation}
I= \int_0^{k_f} dJ(k,\Phi)
\end{equation}

Since our concern is what happens to tunneling electrons that seems to come out of the barrier before entering it, it makes sense to ask if there is
density of states (DOS) under the barrier as well. Again for mesoscopic systems this too can be determined from internal wave-functions as well as
asymptotic wave-function. Note that the internal wave-function in tunneling regime is not consistent with the axioms of quantum mechanics because they
cannot form a wave packet, they are not elements of Hilbert space, they are at most analytic continuation of wave-functions above the barrier.
As usual, the asymptotic wave-function does not care if there is a state inside the ring or not and so if it gives
the same value as that determined from analytic continuation of wave-function inside the ring, it can only mean two things. First, tunneling electrons
do propagate under the barrier and secondly the asymptotic formalism settles a theoretical problem in the sense that analytical
continuation implies physical continuation.
So this is a theoretical experiment in the sense that what cannot be concluded from axioms of quantum mechanics can be concluded from the idea
of a physical clock like Larmor clock.
Of course currents are a better candidate as they can also be meausred in a practical experiment and they do support this point of view as will
be plotted and shown below. Never the less it does not harm to get further confirmation from DOS because after all DOS is a far more valuable
quantity than just the currents. It is linked to all thermodynamic properties and not just currents. Secondly current conservation ensures
certain features that can be settled purely at a level of numbers without caring for a theory.
DOS cannot be directly measured like one can measure current from magnetization, but is a much more complex concept connected to normalization,
renormalization, regularization etc, and if there is a scope to check purely theoretical consistency
with the Larmor clock, that can be exploited by checking the following Eq. 18. 
So now we have the DOS also leading to a current and current can be practically measured and so in our opinion worth studying
in future very specifically. Currents will peak where DOS peaks and quantitative verification can be made.
So we expect
\begin{equation}
\frac{d\theta_R}{edU}=\pi \rho_d^u =
\frac{m}{2 \hbar^2 s}[\frac{|c|^2}{2 s}(1 - e^{-2 s l_2}) + \frac{|d|^2}{2 s} (e^{2 s l_2} - 1) + (c d^* l_2 e^{i \Phi'} + d c^* l_2 e^{-i \Phi'})]
\end{equation}
Here $\rho_d^u$ is the DOS in region II under the barrier and hence the superscript $u$,
LHS giving the expression that can be obtained from asymptotic wave-function, while RHS is that obtained by
integrating the LDOS from the analytic continuation of
internal wave-function. 
Below we will numerically check and verify this equality to authenticate that the analytic continuation of
internal wave-function
in region II is connected with the probability of tunneling particles. It is to be noted that the LHS, that is $\frac{d\theta_R}{edU}$ as an
expression of DOS do not use the tunneling wave-functions in any way and is a derivation that is independent of whether the states are tunneling
or propagating in region II. It only uses the analyticity of the scattering matrix.
The RHS expression consisting of 3 terms in first brackets, are completely different for evanescent modes than that for propagating modes
and has to be found by integrating the absolute value of wave-function in region II with appropriate normalization constant.
Interestingly, the RHS of Eq. 18 giving DOS under the barrier has some negative terms
signifying there are partial states under the barrier for which counting in real numbers (measure) can yield a negative answer.
But total DOS or $\rho_d^u$ is positive definite because $s=(\frac{2m}{\hbar^2}(U-E))^{1/2}$ is positive. As a result $[e^{2 s l_2} - 1] \geq 0$
and $[1 - e^{-2 s l_2}]  \geq 0$.
Although the negative terms do not dominate over
the positive terms, they do signify presence of processes back in time. However, there is no evidence that such processes can carry signal or information
because for that we need propagating wave-packets. 

One may further say that there are states under the barrier given by Eq. 18 consisting of three terms in first brackets. But all three terms are not
responsible for carrying current. Only the last term carries current as Eq. 14 confirms this. An expression like $e^{2sl_2}$ on RHS can grow
indefinitely as $l_2$ is increased and to balance that the current carrying part of DOS that is $(c d^* l_2 e^{i \Phi'} + d c^* l_2 e^{-i \Phi'})$
become fewer and fewer. Current propagates only through these partial states. So if these partial states become fewer and fewer, then the
electron takes lesser and lesser time to traverse the barrier. This shows up in the Hartman effect but as these states do not constitute a
propagating wave-packet they cannot transmit a signal.

\begin{figure}[h!]
\centering
\includegraphics[scale=0.50]{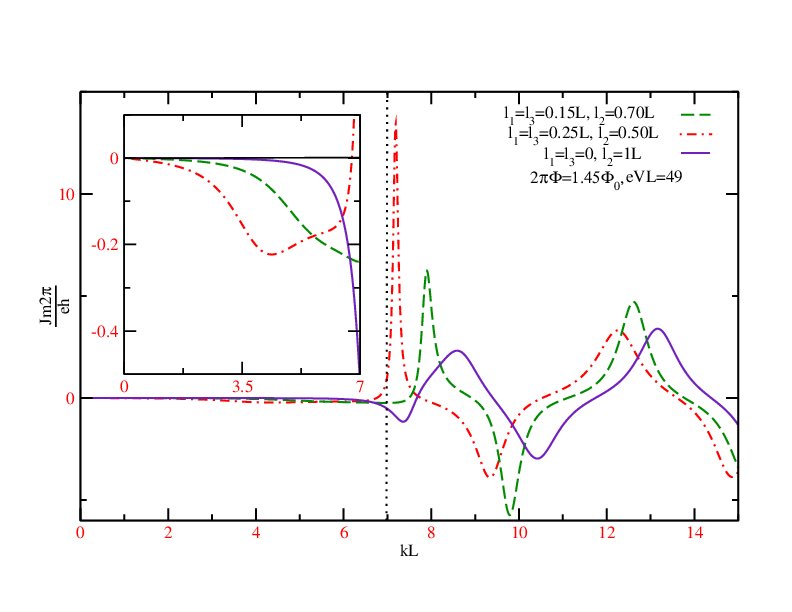}
\captionsetup{labelformat=empty}
\caption{Fig. 3 Plot of current density in region I as a function of wave vector $kL$ as obtained from Eq. 14 for three different systems corresponding
to different lengths of regions I, II and III for the system shown in Fig. 2. Current conservation implies that the current in region II and III will be the same.
Current in region II has to be obtained from Eq. 13 and it confirms this. The vertical dotted line at $kL$=7 give the crossover point for the parameters
used in this figure wherein modes in region II change from evanescent to propagating. We have taken $\hbar$=1 and $2m$=1.}
\end{figure}
\begin{figure}[h!]
\centering
\includegraphics[height=150mm,width=150mm,angle=270]{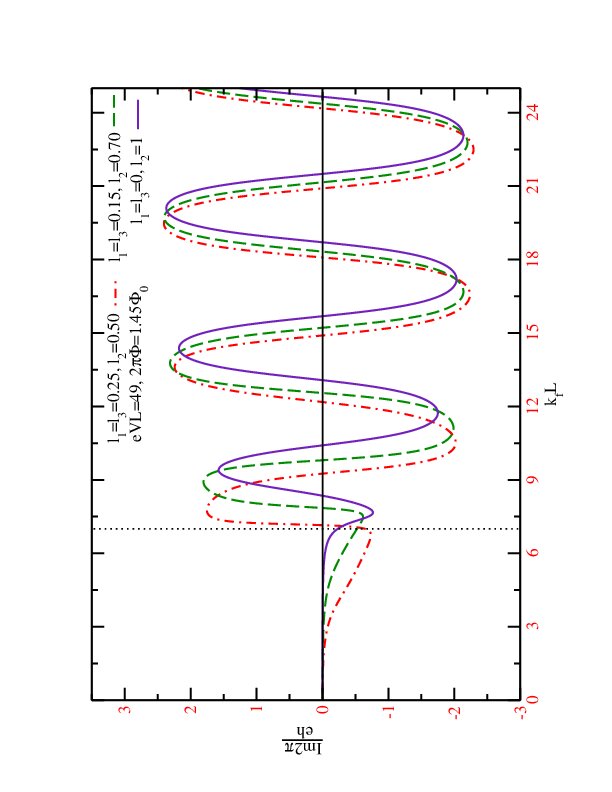}
\captionsetup{labelformat=empty}
\caption{Fig. 4 Plot of measurable current in region I as a function of wave vector $kL$ as obtained from Eq. 16 for the same three different systems
considered in Fig. 3. We have taken $\hbar$=1 and $2m$=1.}
\end{figure}

\section{Results and discussions}

We therefore proceed to demonstrate that Eq. 14 give the same current as Eq. 16 and that the equality claimed in Eq. 18 is true.
One can set up the boundary conditions in Eqs. 7 and 8
that lead to a set of simultaneous equations, that can be solved to obtain
$R$, $a$, $b$, $c$, $d$, $f$, and $g$ 
in terms of known parameters like $l_1$, $l_2$, $l_3$, incident wave vector $k$ and applied magnetic field
through the parameters $\alpha$, $\beta$ and $\gamma$.
In Fig. 3 we plot the current density versus $kL$ which by current conservation has to be the same throughout the ring.
For evanescent regimes we use Eq. 14 and for propagating Eq. 15, respectively.
Three sets of parameter values has been used and mentioned inside the figure.
Note that region II has a positive constant potential of $eUL$=49. That means for $k$ values up to $kL$=7.0 the region II carry only
evanescent states and so a vertical dotted line has been drawn at $kL$=7.0 as a guide to the eye.
The solid curve correspond to a situation wherein $l_1=l_3=0$ implying the entire ring has a positive potential $eUL=49$.
Which means for $kL$ values up to $kL=7$, the entire ring carries evanescent modes only.
The inset to Fig. 3 accentuates this range from $kL=0$ to $kL$=7.0.
Above $kL=7$ the states become propagating.
The dashed curve and dash-dotted curve correspond to situations
when the states in the ring are partially propagating and partially evanescent, wherein regions I and III are propagating while
region II (see Fig. 2) will have evanescent modes up to $kL=7$. Above $kL=7$ the entire ring will carry propagating modes only.
This current density is identically the same as calculated from the alternate formalism of Eq. 16 for all these situations,
which means also for the situation of our interest that is the solid curve of Fig. 3 which confirms what we wanted to show.
Our expression for current in Eq. 14 must be correct as it agrees with Eq. 15 which is a formalism that does not require us to treat
the evanescent modes in any special way.

Another interesting feature to be noted from Fig. 3
is that as the region II of Fig. 2, or the region C of Fig. 1 is enlarged by making $l_1$ and $l_3$ shrink to zero keeping total length $L$ unchanged,
then for energy $E<eU$ current in the ring make a gradual shift from paramagnetic to diamagnetic.
This is shown in the inset to Fig. 3.
When the length
of region II is $l_2=0.5L$, then the current in the ring change from diamagnetic to paramagnetic at $kL=6.68$ (dash-dotted curve).
When $l_2$ is increased to $0.7L$ then this crossover happens at $kL=7.47$ (dashed curve).
For the solid curve, this cross over occur at $kL=7.67$. This diamagnetic to paramagnetic oscillation in current is a property
of propagating modes only, and when the entire ring is in the tunneling regime then under the barrier current can only be diamagnetic.
This has to do with the fact that propagating modes have parity effect wherein alternate states alter in parity. Even states have an even parity
with an even number of nodes while odd states have odd parity with odd number of nodes. Evanescent mode wave-function is not wave like in nature
at all. They have no nodes and so also no parity.
One may ask the question if there is an observable diamagnetic shift of the currents due to a partial or complete tunneling region inside the ring.
Observable current given by Eq. 17 
will be the area under the curves in Fig. 3 up to a Fermi wave-vector $k_fL$. In Fig. 4 we have plotted this integrated current
as a function of $k_fL$ and at low energies we do see a diamagnetic shift up to $k_fL=8.35$. But there is no such discernible diamagnetic shift
at higher energies. But one can always have higher energy evanescent modes in the ring that show a diamagnetic shift at such energies.
Such high energy evanescent modes are possible in presence of sub-bands due to lateral confinement as higher sub-bands have higher propagating thresholds.
Interestingly this integrated current show roughly
same magnitude of current in the evanescent regime as that above the barrier. The peak values of the current at the first peak is roughly -1 while
the propagating current peak value goes to $\pm 2$ (see Fig. 4). This is because in presence of partial or full tunneling the current is small in magnitude but
always diamagnetic resulting in a larger area under the curves. Propagating state current alters between paramagnetic and diamagnetic and has a lot
of cancellation effect.

For completion we have also plotted DOS for the region II
in Fig. 5. We plot the RHS and the LHS of Eq. 18 that coincide with each other for all three graphs as expected. Curves are for the same value
of parameters used in Fig. 3 and the same convention is followed.
Note that all three curves show that when DOS under the barrier is high then current is also high.

\begin{figure}[h!]
\centering
\includegraphics[height=150mm,width=150mm,angle=0]{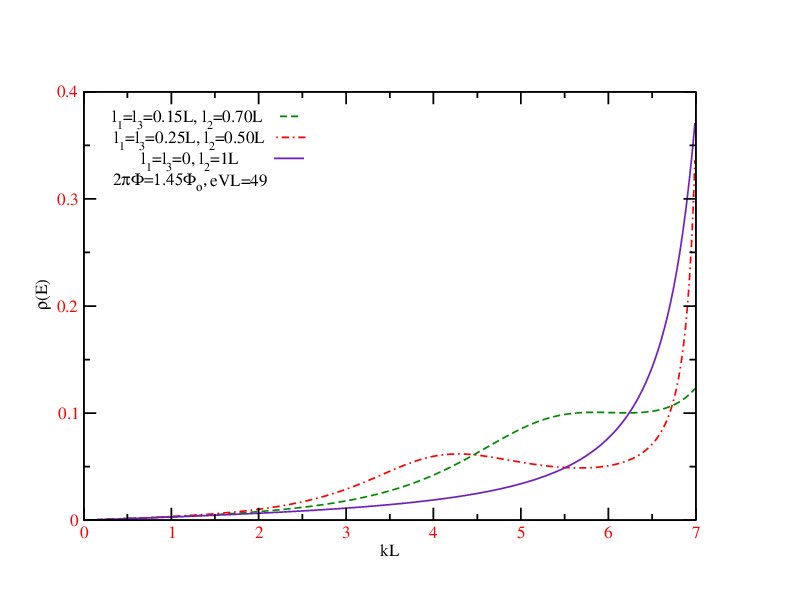}
\captionsetup{labelformat=empty}
\caption{Fig. 5 Plot of DOS in region II only for the same set of parameter values used in Fig. 3 and Fig. 4. We have taken $\hbar$=1 and $2m$=1.}
\end{figure}

\section{Conclusion}
In summary, asymptotic theory works for current as well as DOS and has been also used to show that propagation under the barrier is a reality.
Propagation under the barrier is not a well defined problem in quantum mechanics and so the asymptotic theory that goes beyond quantum
mechanics, uses the idea of physical clocks and analyticity of scattering matrix elements help confirm this. This is referred to as
a theoretical experiment but one can also make clean practical experiments to confirm this further. It also
provides evidence that time reversed states have a role to play in propagation under the barrier. So there is a lot of merit in trying
to address the question of propagation time and superluminality of tunneling particles.

\clearpage

\bibliographystyle{References}

\end{document}